\documentclass[prb,twocolumn]{revtex4-2}
\usepackage{graphicx}
\usepackage{color}
\begin{document}
\title{How to compute density fluctuations at the nanoscale}
\author{Peter Kr\"{u}ger}
\affiliation{
  Materials Science Department,
  Graduate School of Engineering,
  Chiba University, Chiba 263-8522, Japan}  
\begin{abstract}
The standard definition of particle number
fluctuations based on point-like particles
neglects the excluded volume effect. This leads to a large and systematic
finite-size scaling and an unphysical surface term in the isothermal
compressibility.
We correct these errors by introducing a modified pair distribution
function which takes account of the finite size of the particles.
For the hard sphere fluid in one-dimension, we show that the
compressibility is strictly size-independent and we reproduce this
result from the number fluctuations calculated with the new theory.
In general, the present method eliminates the leading finite-size effect,
which makes it possible to compute density fluctuations accurately in
very small sampling volumes, comparable to the single particle size.
These findings open the way for obtaining the local compressibility
from fluctuation theory at the nanometer scale.
\end{abstract}

\maketitle

\section{Introduction}
From fundamental principles of statistical mechanics,
response function are directly linked to
the fluctuations of some extensive variable,
e.g. the specific heat is related to energy fluctuations
and the compressibility to volume or number fluctuations~\cite{landau}.
In inhomogeneous matter, intensive variables, including response functions,
are position dependent.
This raises the question whether fluctuation theory can still be used
at length scales comparable to the particle size.

Various methods have been proposed for generalizing thermodynamics to
systems at the nanometer scale~\cite{hill62,hillbook,carrete08,schnell11,chamberlin15,miguel17,dong23,dong23b},
many of which are based on Hill's thermodynamics
of small system~\cite{hillbook}.
Hill considered statistical ensembles of finite-sized systems
such as droplets and clusters. Here we focus on the related but somewhat
different problem of a finite, open subvolume inside a large system.
We ask the question whether
the response functions
can be accurately calculated from the knowledge of the fluctuations
inside the subvolume alone.
This is a crucial issue for confined or otherwise inhomogeneous systems,
where the local
intensive variables, including the response functions, may vary in space at the nanometer scale.
For example, the spatial variation of the solvent density fluctuations around a solute molecule,
has been shown to give important insights into the thermodynamics of solvation~\cite{rajamani05}
and about drying and wetting phenomena at surfaces and large solute particles~\cite{evans17,coe23}.
In order to discuss such systems
using thermodynamic language, the response functions of a small open
subvolume must be unambiguously defined and computed in an efficient way.

Building on Hill's approach, a theory of nanothermodynamics has been formulated
over the last two decades by the Trondheim group~\cite{dickbook}.
In their small system method, response functions  
are obtained from the fluctuations inside a finite subvolume
embedded in a large reservoir~\cite{schnell11}.
It was found that in homogeneous fluids,
the particle number fluctuations and the isothermal compressibility
are strongly size-dependent at the
nanometer scale~\cite{strom17}. For example,
the compressibility of liquid water in a subvolume of size 1~nm$^3$, 
was found to be three times larger than the bulk compressibility.
This was attributed to a large surface term.

Here we argue that such a large surface term in the compressibility
of a fluid is a numerical artefact induced by the discreteness of the
particle number operator. We show that the surface term essentially
vanishes when the excluded volume effect is
taken into account in the definition of the density fluctuations.
The importance of the excluded volume for the finite-size
effects of the X-ray scattering factor has been stressed
recently~\cite{dohn23}.
Here we demonstrate that the isothermal
compressibility of a hard sphere fluid (HSF) in one dimension (1$D$)
is size-independent down to the smallest physically meaningful size of a
single particle. We show that this result can also be obtained
from the particle number fluctuations in an open subvolume
if the compressibility equation is corrected in two ways.
First, a modified radial distribution function (RDF) is introduced
which takes account of the finite particle size, i.e. the excluded volume
effect. The modified RDF can be obtained from the usual RDF
by a simple convolution.
Second, the compressibility equation must be evaluated
using the finite-volume Kirkwood-Buff integral (KBI)
theory~\cite{kruger13,dawass18,kruger18,dawass19,simon22}
rather than the standard running KBI~\cite{kb52,bennaim}.
For the 1$D$~HSF at high filling,
we find that the fluctuations obtained
with the modified RDF are virtually 
size-independent down to subvolumes containing a single particle.
At low filling some size-dependence remains, but it can be
made to vanish by using a reduced particle size in the modified RDF.
By demanding that the compressibility be exactly size-independent,
which holds for the 1$D$ HSF in the canonical and the isobaric ensemble,
we define an effective particle diameter
which agrees with the hard-sphere diameter at large filling.
We also apply the method to the HSF in 3$D$ using the RDF in the
Percus-Yevick approximation~\cite{percus} and to the hard sphere solid
and we obtain qualitatively the same results as in 1$D$.
Our findings show that density fluctuations can be defined unambiguously
at the nanometer scale and that the local compressibility
can be computed accurately with sampling volumes of minimum size. 
We note however that for response functions which are directly related to the discrete particle
{\em number}\/ fluctuations rather than the continuous {\em density}\/ fluctuations,
such as the chemical potential derivatives~\cite{schnell11}
and excess free energies in solvation theory~\cite{pratt92}, the present method
might not be suitable and the usual definition of the RDF with point-like
particles~\cite{bennaim} should be kept.
\section{The isothermal compressibility from density fluctuations in different ensembles}
The isothermal compressibility of a piece of matter held at temperature $T$ and
pressure $p$ and occupying a volume V, is a mechanical response function, defined as
$\kappa_T = -(1/V)(dV/dp)_T$. For a system containing $N$ identical particles, the
thermodynamic definition becomes
\begin{equation}\label{kadef}
\kappa_T= -\frac{1}{V} \left( \frac{\partial V}{\partial p}\right)_{NT}
\end{equation}
From fluctuation theory we have
\begin{equation}
\label{kaNpT}
\kappa_T= \left.\frac{1}{k_BT}\frac{\delta V^2}{V}\right|_{NT}
= \left.\frac{V}{k_BT}\frac{\delta \rho^2}{\rho^2}\right|_{NT} 
\end{equation}
where $\rho=N/V$ is the particle number density.
We write $A\equiv \langle A\rangle$ for the average
and $\delta A^2 \equiv \langle A^2\rangle-\langle A\rangle^2$
for the fluctuation of a quantity~$A$.
In Eq.~(\ref{kaNpT}) the statistical averages are taken in
the isothermal-isobaric ($NpT$) ensemble.
In fluids, it is often more convenient to work in the grand-canonical
($\mu VT$) ensemble
and the following definition is commonly used, which
relates the compressibility to the particle number
fluctuations in an open volume~$V$ 
\begin{equation}\label{kamuVT}
{\kappa}_{TV}
= \left.\frac{V}{k_BT}\frac{\delta \rho^2}{\rho^2}\right|_{VT}
= \left.\frac{1}{\rho k_BT} \frac{\delta N^2}{N}\right|_{VT} \;\;.
\end{equation}
Clearly, the only difference between Eq.~(\ref{kaNpT}) and Eq.~(\ref{kamuVT})
is the statistical ensemble, which is indicated
in $\kappa_{TV}$ by the extra subscript~$V$.
For a macroscopic system, all ensembles are equivalent and so
${\kappa}_{TV} =\kappa_T$.
For small systems there is no general equivalence between ensembles.
Eq.~(\ref{kaNpT}) is valid on general statistical mechanical grounds
for any number of particles~$N>0$.
The only condition is that that the
{\em reservoir}\/ is large enough such that the
volume fluctuations are approximately gaussian.
For the HSF in 1D, we shall prove explicitly that
Eq.~(\ref{kaNpT}) holds for any~$N$.
However we find ${\kappa}_{TV}\ne \kappa_T$ and
thus Eq.~(\ref{kamuVT}) should not be used to 
calculate the compressibility of small systems.
The reason why ${\kappa}_{TV}\ne \kappa_T$ can be understood as follows.
The compressibility $\kappa_T$
is obviously a continuous variable as are all the quantities
used in Eqs~(\ref{kadef},\ref{kaNpT}). In contrast,
the particle number $N$, whose fluctuations are used in Eq.~(\ref{kamuVT})
is a discrete variable. For nanosystems, $\delta N^2$ is of the order of~1,
and the probability distribution of $N$ is not at all gaussian.
As a consequence
$\delta N^2/N^2|_{VT}\equiv \delta \rho^2/\rho^2|_{VT}$ differs from
$\delta \rho^2/\rho^2|_{NT}$.
However, Eq.~(\ref{kamuVT}) is more convenient for fluids than
Eq.~(\ref{kaNpT}) because Eq.~(\ref{kamuVT}) corresponds to an open
volume in an $\mu VT$ ensemble.
Using a finite open volume, one can, in principle, study the
compressibility, density and concentration fluctuations {\it locally}\/  
in inhomogeneous or confined fluids~\cite{dickbook}.
This is not possible with Eq.~(\ref{kaNpT}) which corresponds
to a closed system.

The fact that $N$ is a discrete variable is related to the
assumption of point-like particles, which is implicit in the definition
of the $n$-particle densities. The 1-particle density is
\begin{equation}\label{rho1}
\rho({\bf r})=\langle \sum_{i}\delta({\bf r}-{\bf r}_{i})\rangle
\end{equation}
where the brackets denote the ensemble average and $i$ counts the particles.
We define the 2-particle density as
\begin{equation}\label{R2}
R({\bf r}, {\bf r}')
=\langle \sum_{i,j}
\delta({\bf r}-{\bf r}_{i})\delta({\bf r}'-{\bf r}_{j})\rangle \;.
\end{equation}
Note that the diagonal terms $i=j$ are included in the sum.
In the more common definition~\cite{bennaim}, denoted $\rho^{(2)}$,
the diagonal terms are omitted.
The two definitions are equivalent and 
we have $R=\rho^{(2)}+\delta({\bf r}-{\bf r}')\rho^{(1)}$.
We use $R$ because it is a more suitable starting point 
for the density fluctuations for particles of finite size,
which will be introduced below.
According to Eq.~(\ref{rho1})
the {\em instantaneous}\/ particle density for a single microstate
diverges at the particle positions ${\bf r}_i(t)$.
While this may approximately be true for the mass density in a gas of atoms,
it is not a good model for the present purpose, because it does not
account for the size of the particles, which may be
atoms or molecules.
Instead of the mass density, it is better to think of the electronic
density which determines the size of the particles and which
explains the excluded volume effect by Pauli repulsion.
If the individual particles are given a finite size and density, then
$N$ becomes a continuous variable,
because a particle near the boundary of an open volume~$V$,
can be partially inside and partially outside of~$V$.
In section~\ref{modpdf}
we shall introduce modified particle densities
which take into account the finite size of the particles.
We shall show that the error of ${\kappa}_{TV}$ can thereby be corrected for.

The particle number fluctuations~$\delta N^2$
in an open volume~$V$ are given by
\begin{equation}\label{dN2}
\delta N^2 = \int_V d{\bf r}_1\int_V d{\bf r}_2
R({\bf r}_1, {\bf r}_2) - \left(\int_V d{\bf r}\rho({\bf r})\right)^2 \;.
\end{equation}
In the following we focus on a homogeneous and isotropic fluid,
where $\rho$ is a constant and $R$ only depends on the
distance $r=|{\bf r}_1-{\bf r}_2|$.
Then Eq.~(\ref{dN2}) can be simplified to~\cite{kruger13}
\begin{equation}\label{dN2R}
\delta {N}^2 = V\int_0^{L} ({R}(r) -\rho^2)w(r)dr
\end{equation}
where $w(r)$ is a geometrical weight function
that depends only on the size and shape
of the volume~$V$~\cite{kruger13,kruger18} and $L$ is the maximum distance
in~$V$.
For a 1$D$ system of length $L$ we have $w(r)= 2 (1-r/L)$ and
for a 3$D$ sphere of diameter $L$ we have
$w(r)=4\pi r^2 [1-3r/(2L)+(r/L)^3/2]$.
%
%
Equation~(\ref{dN2R}) can be written in terms of the RDF,
which is given by 
\begin{equation}\label{gR}
g(r)=[R(r)-\rho\delta({\bf r})]/\rho^2
\end{equation}
With $N\equiv\langle N\rangle=\rho V$, we obtain,
from Eqs.~(\ref{dN2R},\ref{gR})
for the number fluctuations per particle~\cite{kruger13}
\begin{equation}\label{nu2R}
\nu \equiv  \frac{\delta N^2}{\langle N\rangle} 
= 1+{\rho}\int_0^{L} (g(r) -1)w(r)dr \;.
\end{equation}

\section{Hard sphere fluid in 1$D$}
\begin{figure}[tbp]
\begin{center}
\includegraphics[width=0.75\columnwidth]{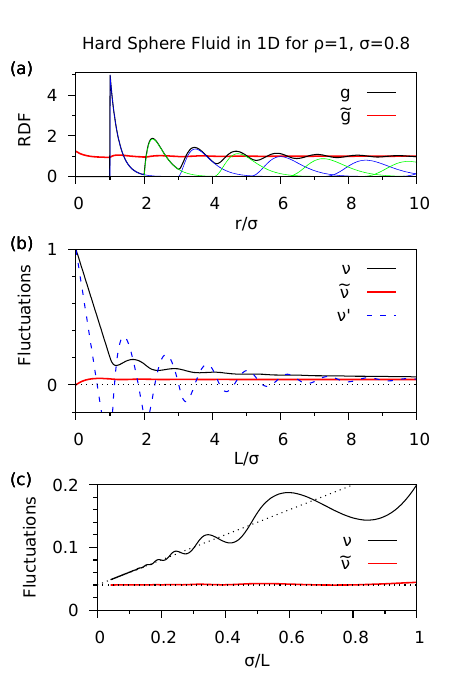}
\end{center}
\caption{
Results for hard sphere fluid in $1D$ with $\rho=1$, $a=\sigma=0.8$.
(a) Usual RDF $g(r)$ (black line) with components $P_n(r)$, Eq.~(\ref{pnr})
for $n$ even (green) and odd (blue).
Modified RDF $\tilde{g}(r)$ (red).
(b) Relative number fluctuations $\nu$ and ${\tilde\nu}$
($=\delta N^2/\langle N\rangle$)
 as a function of finite system size $L$.
 $\nu'$ (green) is obtained with $g(r)$ and the ``Running-KBI'',
 i.e. by putting $w(r)=2$ in Eq.~(\ref{nu2R}).
(c) $\nu$ and ${\tilde\nu}$  as a function
of $1/L$. The thin black lines are linear interpolations, namely
$\nu=0.04$ and $\nu=0.04+0.2\sigma/L$.
}\label{fig1}
\end{figure}

We first consider the hard sphere fluid (HSF) in 1$D$, also known as the
Tonks model~\cite{tonks}, and we study the size dependence of the
isothermal compressibility and of the density fluctuations. 
In 1$D$, force~$F$ and pressure $p$ are the same.
The system length~$L$ is the 1$D$ hypervolume~$V$.
With these notations, the exact equation of the state of the Tonks
model~\cite{tonks} with $N$ particles of diameter $\sigma$ is
given by $V\equiv L=N(\sigma+k_BT/p)$.
In the canonical ensemble,
this equation of state is valid for any number of particles~$N$ and
any system size~$L>N\sigma$.
The isothermal compressibility follows directly from the
definition, Eq.~(\ref{kaNpT}) and we obtain
\begin{equation}\label{kato1}
\kappa_T=\frac{(1-\rho\sigma)^2}{\rho k_BT} \;.
\end{equation}
Note that the compressibility~(\ref{kato1}) is clearly size-independent.
Next we calculate~$\kappa_T$ from fluctuation theory.
First we consider volume fluctuations in the $NpT$ ensemble.
To this end, we employ the grand-canonical RDF of the Tonks model,
which is known analytically~\cite{salzburg53} 
\begin{equation}\label{gofr1D}
g(r) = \frac{1}{\rho}\sum_{n=1}^\infty P_n(r) \;,
\end{equation}
where
\begin{equation}\label{pnr}
P_n(r)=\theta(x) \lambda^n \exp(-\lambda x) {x^{n-1}}/{(n-1)!}
\end{equation}
Here we have put $x = r-n\sigma$, $\lambda=\rho/(1-\rho\sigma)$
and $\theta(x)$ is the Heaviside step function.
$P_n(r)dr$ is the probability that the distance between
a given particle and its $n$-th nearest neighbor to the right,
lies in $(r,r+dr)$.
The RDF $g(r)$ and its components $P_n(r)$ are plotted in Fig.~\ref{fig1}~(a)
for a HSF in 1$D$ with $\rho=1$, $\sigma=0.8$.
In the 1$D$ HSF, the distance distribution $P_N(r)$
can also be interpreted as the length distribution
of a closed system of $N$ particles in a $NpT$ ensemble.
This was explicitly shown by Tonks~\cite{tonks} for $N=1$ and
the argument can easily be generalized to $N>1$.
It follows that the expectation value and the variance of
the system length ($L$) in the $NpT$ ensemble
can be obtained from the first two moments of the $P_N(L)$ distribution,
i.e. Eq.~(\ref{pnr}) with $n=N$, $r=L$.
The $k$-th moment is $\langle L^k\rangle = \int_0^\infty L^k P_N(L) dL$.
The calculation is straightforward and gives
\begin{equation}\label{dr2r}
\langle L\rangle = N/\rho\;,\quad
{\delta L^2 } = N {(1-\rho\sigma)^2}/\rho^2 \;.
\end{equation}
These expressions are valid for any~$N$, i.e. for any system size in the
$NpT$ ensemble.
Upon identifying $\delta L^2/\langle L\rangle$ with
$\delta V^2/\langle V\rangle$ ,
we obtain, from Eq.~(\ref{kaNpT}), exactly the same expression
for $\kappa_T$ as from the thermodynamic definition Eq.~(\ref{kato1}).
Thus we have demonstrated that for the 1$D$ HSF,
the hyper-volume fluctuations in the $NpT$ ensemble give
the exact, size-independent, isothermal compressibility.

Next we study the density fluctuations of the 1$D$ HSF
in the grand-canonical ensemble, on a open segment of
finite length~$L\equiv V$.
We have $\langle N\rangle=\rho V$ and $\delta N^2$
can be calculated from Eqs~(\ref{nu2R},\ref{gofr1D},\ref{pnr}).
The relative fluctuations $\nu\equiv \delta N^2/\langle N\rangle$
are shown in Fig.~\ref{fig1}~(b,c) (black lines).
$\nu$ has a pronounced
size dependence and becomes linear in $1/L$ for large $L$.
For $L\rightarrow \infty$, $\nu$ goes to $(1-\rho\sigma)^2$
($=0.04$ in this case).
This is the exact limit, which follows from Eq.~(\ref{kato1})
and the fact that $\kappa_{TV}=\kappa_{T}$
in the thermodynamic limit.
Since the exact isothermal compressibility~$\kappa_T$, Eq.~(\ref{kato1}),
is strictly size-independent, it is clear that the large size dependence
of $\nu$ (and thus $\kappa_{TV}$) seen in Fig~\ref{fig1} (b,c),
is a non-physical result.
We conclude that the particle number fluctuations~$\nu$
defined in the usual way with Eqs~(\ref{R2},\ref{dN2}) cannot be used
to compute the isothermal compressibility for finite-size systems.
For completeness, we also show the result obtained
when using the standard or ``running''-KBI instead of the
finite-volume KBI, Eq.~(\ref{nu2R}). For the running KBI,
the infinite-volume expression is simply truncated at the upper bound~$L$.
It can be computed from Eq~(\ref{nu2R}) upon replacing
$w(r)$ by $2$ in 1$D$ or $4\pi r^2$ in $3D$.
The number fluctuations of the 1$D$ HSF, computed with the running KBI
are plotted in Fig.~\ref{fig1}~(b) as $\nu'$ (blue dashed line).
The function $\nu'(L)$ oscillates strongly and thus
has an even larger size dependence than $\nu$.
Most importantly, $\nu'$ becomes negative for certain system sizes $L$,
which is obviously wrong, since fluctuations cannot be negative on
mathematical grounds.
This means that the running-KBI cannot be used to assess finite-volume
density fluctuations.

\section{Modified RDF with excluded volume}
\label{modpdf}
We have seen above that the difference between $\kappa_{TV}$ and
$\kappa_T$ is related to the discreteness of the particle number and
we have shown that the usual, point-like definition of the particle densities
leads to unphysical results for the compressibility~$\kappa_{TV}$
of small systems.
Now we go beyond the point-like particle picture and take account of
the fact that real particles have a finite size, by replacing
the $\delta({\bf r})$ functions
in the definition of the $n$-particle densities, by a regular
function $\Delta({\bf r})$
which describes the finite density of the individual particles.
Here we consider hyper-spherical particles with diameter $a$
and constant internal density,
\begin{equation} \label{Delta}
\Delta(r)=\left\{\begin{array}{ccc}
1/V_0&& (r<a/2)\\
0 && (r>a/2)\\ \end{array}\right.
\end{equation}
where $V_0$ is the hyper volume, i.e. $V_0=a$ in 1$D$ and
$V_0=\pi a^3/6$ in 3$D$.
Such a step-like density profile with $a=\sigma$
is the natural choice for the hard-sphere model.
However, other simple density profiles lead to very similar
results and to the same conclusions,
as we shall show explicitly for a gaussian profile
in a forthcoming paper.

By attributing a finite size to the particles in the modified $n$-particle densities,
the number of particles $N$ in a small volume becomes
a continuous rather than a discrete variable. However, the underlying statistical mechanical calculation are not changed in any way,
and the discreteness of the particle number as a physical observable is respected.
The variable $N$ is taken as continuous only when computing the particle density
fluctuations in a finite open volume from the microscopic states.
The modified quantities, obtained upon replacing
$\delta({\bf r})$ by $\Delta({\bf r})$,
are denoted by ${\tilde\rho}$, ${\tilde R}$, ${\tilde \nu}$ etc.
Introducing a spherically symmetric internal density $\Delta(r)$
does not break the symmetry of the fluid. As a consequence,
the modified 1-particle density is unchanged,
${\tilde\rho}=\rho$ and the modified 2-particle density, given by
\begin{equation} \label{Rtrr}
{\tilde R}({\bf r}, {\bf r}')
=\langle \sum_{ij}
\Delta({\bf r}-{\bf r}_{i})\Delta({\bf r}'-{\bf r}_{j})\rangle \;,
\end{equation}
only depends on the particle distance~$|{\bf r}-{\bf r}'|$.
The function ${\tilde R}(r)$
can be obtained from $R(r)$ by a convolution
\begin{equation}\label{rchir}
{\tilde R}(r)=\int \chi(r,r') R(r')dr' \;,
\end{equation}
as we have shown previously~\cite{miyaji21}. Note that in Ref.~\cite{miyaji21},
the convolution was derived for ${\tilde g}(r)$ from $g(r)$,
where the diagonal terms ($i$=$j$) are omitted, but this has no
effect on the convolution, i.e. the functions $\chi(r,r')$
for $R\rightarrow{\tilde R}$ are the same as
those for $g\rightarrow{\tilde g}$ in Ref.~\cite{miyaji21}.
It is easy to see that in 1$D$ we have
$\chi=(a-|r-r'|)/a^2$ for $|r-r'|<a$ and $\chi=0$ otherwise.

The particle number fluctuations can be computed in the same way
as before upon replacing $R(r)$ by ${\tilde R}(r)$
in Eq~(\ref{dN2R}).
We define the modified RDF~${\tilde g}(r)$ as 
\begin{equation}\label{gt}
{\tilde g}(r)={\tilde R}/\rho^2 \;.
\end{equation}
In this definition, the diagonal ($i=j$) terms
in the two-point density Eq.~(\ref{Rtrr}) are included.
In the definition of the usual RDF $g(r)$,
the diagonal terms are omitted because they would lead to a delta-function
at $r=0$. With the modified RDF,
the relative fluctuations are given by
\begin{equation}\label{nut2R}
{\tilde \nu}= {\rho}\int_0^{L} ({\tilde g}(r) -1)w(r)dr \;.
\end{equation}
Note that the constant~1 in the expression for $\nu$, Eq~(\ref{nu2R}),
which compensates for the omitted diagonal terms in $g(r)$,
is absent in Eq~(\ref{nut2R}).

For the 1$D$ HSF with $\rho\sigma=0.8$, the modified RDF
${\tilde g}(r)$ is shown in Fig.~\ref{fig1}~(a) as the red line.
The large oscillations of $g(r)$ are almost completely suppressed
and ${\tilde g}(r)\approx 1$ for all~$r>\sigma$.
The corresponding particle number fluctuations ${\tilde\nu}$
in an open system of length~$L$ are shown
as red lines in Fig.~\ref{fig1}~(b,c).
Except of $L<\sigma$, ${\tilde\nu}(L)$ is almost constant at~0.04,
which is the correct value in the thermodynamic limit.
This means that the density fluctuations are virtually
size-independent down to the smallest physically meaningful
system size of one particle ($L=1/\rho$)
in agreement with the analysis in the $NpT$ ensemble.
Importantly, it follows that
$\kappa_{TV}(L)\approx\kappa_T$ $\forall L>1/\rho$,
i.e. the true compressibility~$\kappa_T$
can be approximated, with very good accuracy by computing~$\kappa_{VT}$
from the particle number fluctuations in a open volume~$V$ of
minimum size, provided that the modified RDF is used.

\begin{figure}[tbp]
\begin{center}
\includegraphics[width=\columnwidth]{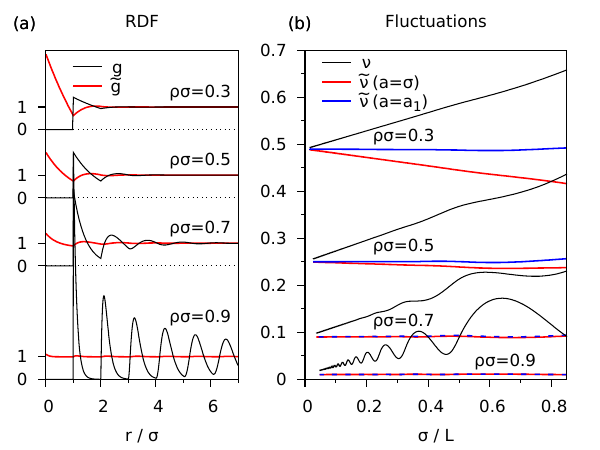}
\end{center}
\caption{Results of hard sphere fluid in 1$D$ as a function of filling fraction $\rho\sigma$.
(a) Usual RDF ($g$) and modified RDF (${\tilde g}$) for $a=\sigma$.
(b) Corresponding number fluctuations.
}\label{fig2}
\end{figure}

In Fig.~\ref{fig1}
we have considered a large filling fraction $\rho\sigma=0.8$,
appropriate for modeling a liquid.
The dependence of the RDF
and the fluctuations on the density $\rho$
is shown in Fig.~\ref{fig2} (a,b).
For decreasing $\rho$, the oscillations of the usual RDF $g(r)$
become smaller and $g(r)$ approaches a step function for $\rho\rightarrow 0$.
At the same time, the modified RDF ${\tilde g}(r)$ deviates more
strongly from~1 for small~$r$ up to about $1/\rho$.
The peak seen at $r=0$ (for small $\rho$) 
is due to the diagonal ($i$=$j$) terms in Eq.~(\ref{Rtrr}).
It evolves to a delta-function in the limit $\rho\sigma\rightarrow 0$.

Looking at Fig.~\ref{fig2} (b), we see that in all cases,
the relative fluctuations tend to the exact result $(1-\rho\sigma)^2$
in the limit $L\rightarrow\infty$.
The functions $\nu(L)$ (black lines)
which were obtained with the usual RDF $g(r)$,
show a strong size dependence that is linear in $1/L$ for large~$L$.
Interestingly, the slope, which may be interpreted as a
surface contribution~\cite{strom17,dawass20}, hardly depends on the
density~$\rho$ (at constant~$\sigma$) but scales linearly with
$\sigma$ (at constant~$\rho$).
This suggests that the surface contribution to $\nu$ is essentially
determined by the size of the individual particles,
or their excluded volume, rather than by long-range correlations
in the fluid.

The fluctuations obtained with the modified RDF are shown
in Fig.~\ref{fig2}~(b) for
two choices of the diameter~$a$ in Eq.~(\ref{Delta}),
namely the ``natural'' choice $a=\sigma$ and a ``corrected'' value
$a=a_1$ defined below.
Considering first $a=\sigma$ (red lines), we see that for high
density ($\rho\sigma > 0.6$, typical for a liquid phase)
the fluctuations ${\tilde\nu}(L)$ are virtually independent of~$L$.
However, for lower densities ($\rho\sigma < 0.6$)
a non-negligible size dependence appears and ${\tilde\nu}(L)$ deviates
from ${\tilde\nu}(\infty)$ by a negative term which is linear in $1/L$.
We recall that the aim of the RDF modification is to obtain
density fluctuations in finite open systems that agree
with those in the $NpT$ ensemble
and which thus yield the physically correct compressibility,
i.e. $\kappa_{TV}=\kappa_T$.
Since the $\kappa_T$ is strictly size-independent in the 1$D$ HSF,  
we choose the parameter~$a$ in Eq.~(\ref{Delta})
such that the size-dependence of the number fluctuations
${\tilde\nu}(L)$ becomes as small as possible.
Empirically, we find that $a=a_1$, where
\begin{equation}\label{a1}
a_1/\sigma =1-(1-\rho\sigma)^3 \;,
\end{equation}
gives excellent results for all densities, as seen in Fig.~\ref{fig2}~(b),
blue lines.
Note that the difference between $a_1$ and the hard-sphere diameter $\sigma$
is negligible for high density,
e.g. for $\rho\sigma=0.8$ it is merely 0.8\%.
In the low density limit, $\rho\rightarrow 0$, we have
$a_1/\sigma\rightarrow 0$,
which means that the modified RDF goes back to the usual,
point-like definition. This is consistent with the fact
that in the zero density limit, the HSF behaves like an ideal gas.

The need for reducing the particle size may seem to introduce some
empirical element into the theory. This is not the case, however,
since the reduction factor is determined by a physical condition.
Indeed, our aim is to define the number fluctuations in a small volume
in such a way as to obtain the same density fluctuations as in the isobaric
ensemble, or equivalently we demand $\kappa_{TV}=\kappa_{T}$
in Eqs~(\ref{kaNpT},\ref{kamuVT}).
Since $\kappa_T$ is size-independent as shown above, so must be $\kappa_{TV}$
and ${\tilde\nu}$,
which is the condition we use for the particle size reduction.
In the Appendix, we show that, if the compressibility of a small open sub-system of a homogeneous fluid depends on the sub-system size,
then contradictions with general thermodynamic principles arise. A physically meaningful definition of the local compressibility is size-independent.
As a consequence, if the compressibility is to be obtained from fluctuation theory,
then the local density fluctuations in a small open system should be free of any systematic
size-dependence.
We note however, that the density fluctuations are strictly size independent only for the hard-sphere fluid.
In systems with attractive interactions
the particle number fluctuations may display a significant size dependence which is related to the nucleation of
a liquid-vapor interface~\cite{rovere93}.
Application of the present formalism to such cases will be reported elsewhere.
The new method will remove the spurious $1/L$ dependence due
to the neglect of the excluded volume effect, but can be expected to leave the physical size dependence qualitatively unchanged.

\section{Hard sphere fluid in 3$D$}
\begin{figure}
\begin{center}
\includegraphics[width=\columnwidth]{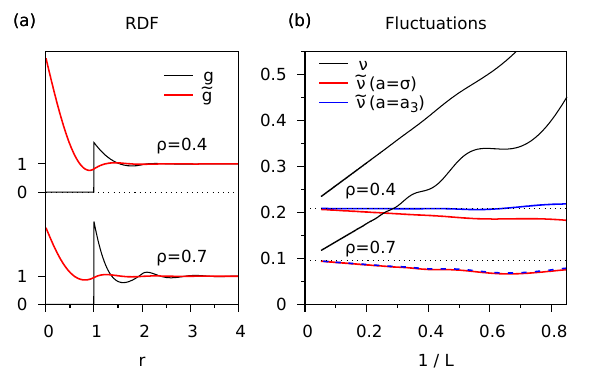}
\end{center}
\caption{Hard sphere fluid in 3$D$ with $\sigma=1$ and
  different densities $\rho$.
  (a) Usual RDF ($g$) and modified RDF (${\tilde g}$) for
  $a=\sigma$.
(b) Corresponding number fluctuations.
}\label{fig3}
\end{figure}

For the HSF in 3$D$ exact analytic solutions are not known.
Here we use the solution of the Ornstein-Zernicke integral equations
in the Percus–Yevick approximation (PYA)~\cite{percus}. 
The corresponding RDF was given by Wertheim~\cite{wertheim64} and
we use the implementation by Kelly et al.~\cite{kelly16}.
In the 3$D$ HSF the density $\rho\equiv N/V$ is related to
the filling fraction~$\eta$ as $\eta = \rho \sigma^3\pi/6$.
We put $\sigma=1$ in the following.
In Fig.~\ref{fig3}~(a),
the RDF $g(r)$ is shown for $\rho=0.4$ and $\rho=0.7$.
As in 1$D$, the modified RDF ${\tilde g}(r)$
can be obtained by a convolution, Eq.~(\ref{rchir}).
In 3$D$, for the constant density profile Eq.~(\ref{Delta}),
the kernel function~$\chi$ is given by~\cite{miyaji21}
\begin{equation}
\chi(x',x)= 6({x'}/{x}) [ 2\xi_2(x',x)-3\xi_3(x',x)+\xi_5(x',x)]
\end{equation}
where $x=r/a$ and
\begin{equation}
\xi_n(x',x)= \left([{\rm min}(1,x+x')]^n - |x-x'|^n\right)/n \;.
\end{equation}
This holds for $|x-x'|<1$ while $\chi=0$ for $|x-x'|>1$.
The modified RDF ${\tilde g}(r)$ with $a=\sigma$ is shown in
Fig.~\ref{fig3}~(a). Compared to $g(r)$, the oscillations are strongly suppressed
in ${\tilde g}(r)$, which is very flat and close to~1 for all $r>1.5$.

We note that for non-spherical particles, the hard-sphere density profile of Eq.~(\ref{Delta}) is not suitable.
It should be replaced by a profile obtained by averaging over all particle orientations.
As a consequence, the kernel function~$\chi(r,r')$ will in general not have any simple
analytic form, but the one-dimensional convolution, Eq.~(\ref{rchir}),
is still valid and the numerical calculation will not become more difficult.

The relative number fluctuations $\delta N^2/N$
are plotted in Fig.~\ref{fig3}~(b) as a function of $1/L$,
where $L$ is the diameter of the spherical subvolume~$V$.
The values $\nu$, obtained with the usual RDF $g(r)$, 
show a $1/L$ scaling with a large slope.
For $\rho=0.7$, the value of $\nu$ at $1/L=0.2$, which corresponds
to a subvolume 
containing $\langle N\rangle=46$ particles,
is about twice as large as the bulk value $\nu(\infty)$.
This finding is in qualitative agreement with the study on
liquid water by Strom et al.~\cite{strom17}.

\begin{table}[t]
  \begin{tabular}{cccccccc}
    $\rho$  & $\nu(\infty)$ & $\tilde{\nu}(\infty)$& $\nu$(EoS) & rel.err & $C$&$\tilde{C}$ &
    $a_3/\sigma$ \\
    \hline
    0.1     & 0.6635        & 0.6635   & 0.6611 & 0.003  & 0.318 &\ 0.001 & 0.429\\
    0.2     & 0.4448        & 0.4449   & 0.4406 & 0.010  & 0.454 &\ 0.000 & 0.674\\
    0.3     & 0.3017        & 0.3019   & 0.2949 & 0.023  & 0.499 &\ 0.001 & 0.814\\
    0.4     & 0.2086        & 0.2088   & 0.1974 & 0.057  & 0.498 & -0.002 & 0.894\\
    0.5     & 0.1494        & 0.1495   & 0.1317 & 0.134  & 0.473 & -0.011 & 0.939\\
    0.6     & 0.1141        & 0.1141   & 0.0872 & 0.308  & 0.438 & -0.027 & 0.965\\
    0.7     & 0.0964        & 0.0965   & 0.0570 & 0.691  & 0.397 & -0.049 & 0.980\\
    \hline
  \end{tabular}
  \caption{Relative number fluctuations $\nu$ of the 3$D$~HSF as a function of density $\rho$.
    Result of the linear fits $\nu(L)=\nu(\infty)+C/L$ and
    ${\tilde\nu}(L)={\tilde\nu}(\infty)+{\tilde C}/L$ (for $a$=$a_3$).
    $\nu$(EoS) corresponds to the Carnaghan-Starling equation of state.
    The relative error (``rel.err'') is defined as $\nu(\infty)/\nu$(EoS)$-1$.
    The values $a_3/\sigma$ were used for $\tilde{\nu}$ and correspond to Eq.~(\ref{a3}).
  }\label{tab1}
\end{table}

The results of a linear fit $\nu(L)=\nu(\infty)+C/L$ in the region
$0<1/L<0.4$ are given in Table~\ref{tab1},
also for other densities not shown in Fig.~\ref{fig3}.
We also listed the values of $\nu$ obtained from
the equation of state (EoS) by Carnaghan and Starling~\cite{carnahan}.
The EoS is given by
$p/(\rho k_BT)\equiv Z(\eta)=(1+\eta+\eta^2-\eta^3)/(1-\eta)^3$
where $\eta=\rho\pi\sigma^3/6$ is the filling fraction.
The isothermal compressibility is readily obtained from the definition, Eq.~(\ref{kadef}), and together with $\nu=\rho k_BT\kappa_{T}$, we have
$\nu=(1-\eta)^4/(1+4\eta+4\eta^2-4\eta^3+\eta^4)$.
Despite its simplicity, the Carnaghan-Starling EoS is known to 
be very reliable for low and moderate densities up to about $\rho=0.9$~\cite{mulero}.

From Table~\ref{tab1} we see that the $\nu(\infty)$
values obtained with the RDF in PYA agree very well with the EoS for
low density, with an error of a few percent
for $\rho\le 0.4$. For higher density, the error rises quickly and
reaches 70\% for $\rho=0.7$. Therefore, the results obtained for
$\rho>0.4$ must be considered preliminary due to the approximate
nature of the RDF in PYA.

In Fig.~\ref{fig3}~(b) the fluctuations ${\tilde\nu}(L)$,
obtained with the modified RDF ${\tilde g}(r)$ with $a=\sigma$ (red lines)
are roughly constant for $L>\sigma$, in stark contrast to the large size
dependence of~$\nu(L)$.
In order to make ${\tilde\nu}(L)$ as size-independent as possible,
we use a reduced particle size in the calculation of the modified RDF.
In 3$D$ we put $a=a_3$, where
\begin{equation}\label{a3}
a_3/\sigma = 1-\exp(-5.6\rho\sigma^3)\;.
\end{equation}
The numerical values $a_3/\sigma$ are given in the last column of Table~\ref{tab1}.
This empirical correction works very well for $\rho\le 0.4$.
${\tilde\nu}(L)$ is virtually size-independent as seen directly from Fig.~\ref{fig3}~(b)
and from the negligible values of the slope $\tilde{C}$.
For $\rho>0.5$, the $\tilde{C}$ values are not negligible and
some $1/L$-like size dependence remains. We do not attempt to correct for this
by adjusting~$a_3$, because we attribute the residual size dependence
to the quite large error of~$\nu(L)$ (see Table~\ref{tab1})
coming from the RDF in PYA, rather than to the functional form of~$a_3$.

In Table~\ref{tab1}, the values ${\tilde\nu}(\infty)$ and
${\nu}(\infty)$ agree almost perfectly.
Let us note that theoretically,
${\tilde\nu}(\infty)={\nu}(\infty)$ holds whatever the choice of $a$.
Indeed, it is easy to see from the definition~(\ref{Rtrr}) that
attributing a finite density to the particles,
does not change the relative number fluctuations in the thermodynamic limit~\cite{miyaji21}.

\begin{figure}
\begin{center}
\includegraphics[width=0.8\columnwidth]{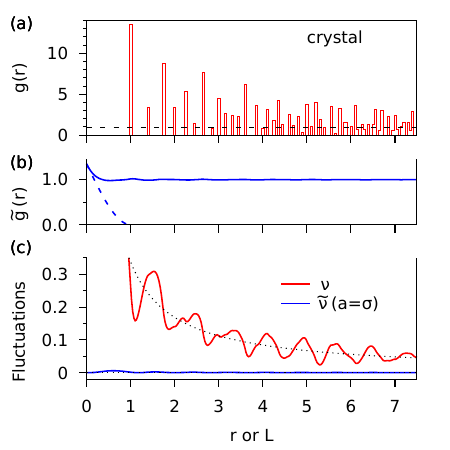}
\end{center}
\caption{(a) RDF $g(r)$ of a perfect face-centered cubic crystal
  The delta functions are replaced by bars of finite width 0.05.
  (b) Modified RDF ${\tilde g}(r)$ with $a=\sigma=(\sqrt{2}/\rho)^{1/3}$.
  The dashed line is the contribution from the diagonal ($i$=$j$) terms
  in Eq.~(\ref{Rtrr}).
  (c) Relative fluctuations $\nu(L)$ (red) and ${\tilde\nu}(L)$ (blue).
  We have $\nu(0)=1$ and $\nu(\infty)=0$.
  The thin dotted line is a guide to the eye given by $f(L)=0.337/L$.
}\label{fig4}
\end{figure}

In this paper we have focused on systems with a single species.
In the case of mixtures, each species $i$ will have a different particle size.
The best choice for the effective particle sizes $a_i$ is not obvious and needs to be studied
in detail. A simple model consists in fixing the ratios $a_i/a_j$ to the corresponding ratios
of the free particle sizes and adjusting the remaining overall parameter
such as to minimize the size dependence of the compressibility, as was done
here for the single species case.

Finally we apply our method to a hard sphere solid at zero temperature.
It has a face-centered cubic crystal structure and the largest
possible filling fraction of $\eta=\sqrt{2}\pi/6$.
The density is $\rho=\sqrt{2}/\sigma^3$.
Since the atoms are immobile, the usual RDF is a sum of delta-functions,
shown as a histogram in Fig.~\ref{fig4}~(a) for $\sigma=1$.
The modified RDF ${\tilde g}(r)$ with $a=\sigma$
is the blue solid line in Fig.~\ref{fig4}~(b).
The blue dashed line is the contribution to ${\tilde g}(r)$
from the terms $i$=$j$ in the two-particle density, Eq.~(\ref{Rtrr}).
These terms are omitted in~$g(r)$, since they correspond to a term
$\rho\delta({\bf r})$ which integrates to~1 in $\nu$ in any system.
For particles with finite density however, the diagonal
terms are non-trivial and their integral leads to a non-singular peak in
${\tilde g}(r)$ near $r=0$ (dashed line in Fig.~\ref{fig4}~b).
This peak fills the ``hole'' of the RDF for $r<\sigma$, 
which is present in all real systems due to the excluded volume effect.
Quite surprisingly, ${\tilde g}(r)$ is virtually constant and equal to~1
for all~$r$, and thus very similar to the RDF of the ideal gas
($g(r)=1$ $\forall r$).
The crucial difference comes from the diagonal terms of the two
particle density that are included in ${\tilde g}$ but not in ${g}$.
In the ideal gas these terms lead to $\nu=1$, while in
${\tilde g}$ of the solid they just fill hole of the excluded volume,
so that ${\tilde \nu}\approx 0$ as seen in Fig.~\ref{fig4}~(c).

The fluctuations are shown in Fig.~\ref{fig4}~(c).
The function $\nu(L)$ has large oscillations, which are due to the long-range
correlations in $g(r)$. It shows the familiar $1/L$ scaling and
converges to the correct thermodynamic limit $\nu(\infty)=0$~\cite{kruger21}.
Note that this trivial result cannot be obtained from the standard
expression of the compressibility equation~\cite{bennaim},
i.e. with $4\pi r^2$ instead of $w(r)$ in Eq.~(\ref{nu2R}).
Indeed, the running-KBI strongly diverges in solids and finite-volume KBI
theory must be used to obtain convergent results~\cite{kruger21,miyaji21}.
While $\nu(\infty)$ is correct, $\nu(L)$ for finite~$L$ is not.
In a perfect crystal at $T$=0~K all atoms are immobile, and so
the density fluctuations and the compressibility are strictly zero,
whatever the size of the system. In other words, the whole functional form
of $\nu(L)$ (i.e. its deviation from zero)
is a finite-size effect and has no physical meaning.
In stark contrast to $\nu(L)$,
the function~${\tilde\nu}(L)$ obtained with the modified RDF is very nearly
constant and equal to zero for any~$L$, which is the
physically correct result.

In this paper we have showed that the modified RDF ${\tilde g}(r)$ should be used instead of the usual RDF $g(r)$
when computing the density fluctuations in finite volumes for obtaining the local compressibility. However, this does not mean that $g(r)$ should be replaced by ${\tilde g}(r)$ in all statistical mechanical relations.
As an example where the modified RDF should not be used, we mention the contact theorem,\cite{roth10}
which holds exactly for the HSF, and which relates
the usual RDF at $r=\sigma$ to the bulk pressure $p$ by $g(\sigma)=p/(k_B T \rho)$.
In this relation, ${\tilde g}(\sigma)$ cannot be used instead of $g(\sigma)$, since ${\tilde g}(r)$
differs from $g(r)$ in principle at all points~$r$, see Fig.~\ref{fig3}(a).
In general, $g(r)$ should only be replaced by ${\tilde g}(r)$
in expressions involving integrals over the RDF.
The infinite volume integral over the RDF is unchanged by the modification~\cite{miyaji21}.
So thermodynamic relations involving integrals over the RDF,
such as the compressibility equation or the Gibbs adsorption theorem~\cite{roth10}, are still valid with
${\tilde g}(\sigma)$ instead of $g(r)$.
For relations involving specific points of the RDF, such as the contact theorem, the usual, point-like RDF $g(r)$ must in general be kept.
A second example where the present method should not be
used is the calculation of the solvation free energy in a hard sphere fluid with the
information theoretical approach~\cite{pratt92,hummer96,garde96}.
In this theory, the probability $P(N=0)$ for finding exactly zero solvent particles in a
volume $V$ that equals the size of the solute particle, is required. By construction,
$P(N)$ is a discrete probability distribution for $N$ particles in $V$ and it can be expressed
in terms of the fluctuations of the number of particle centers inside
the exclusion volume of the solute particle.
For computing $P(N)$, the point-like particle number fluctuations is the natural choice,
while the continuous density fluctuations of extended particles introduced here,
cannot be used directly.
The foregoing examples show that our new approach has some limitations.
The modified RDF should only be used for volume integrals,
and the resulting density fluctuations should not be used to compute
quantities corresponding to discrete values of $N$.

\section{Conclusions}
In summary, we have developed a method for calculating density
fluctuations~$\nu$ and the isothermal compressibility~$\kappa_T$
at the nanoscale, from a modified RDF
which takes account of the finite size of the individual
particles.
This modification removes the large and systematic $1/L$
size-dependence of $\nu$
that is found with the usual method, i.e. with the particles defined as
point-like in the RDF.
We have shown that the $1/L$ term is a non-physical finite-size
effect due to the neglect of the excluded volume.
We have applied the new method to the HSF in 1$D$ and 3$D$.
We find that at high density, typical for
the liquid state, the fluctuations are size-independent
down to the smallest physically meaningful sampling volume 
containing one particle.
For the HSF at low density, some size-dependence of $\nu$ remains,
but it can be made vanish by using a particle diameter in the RDF
which is smaller than the hard sphere diameter.
Simple empirical formula for the optimum size reduction are
given for the HSF in 1$D$ and 3$D$. The method also gives
excellent results for the hard sphere solid.
While we have focused on the HSF in this paper, the method can be
applied to any system, with a suitable definition
of the effective particle size used in the modified RDF.
The main findings should be valid for any fluid,
because they are related to the excluded volume effect,
which is independent of the type of particle interactions.
The new method has major advantages over the
usual method based on point-like particles.
First, numerical calculations of particle number fluctuations
converge much faster as a function of the sampling volume.
For the HSF, a sampling volume containing a single particle
gives already well converged results.
Second, density fluctuations can be computed locally with good accuracy using very small
sampling volumes, thereby minimizing the finite size effects. This should make it possible
to compute the local compressibility and related quantities in inhomogeneous and confined
system, unambiguously and accurately from fluctuation theory. To this end, the RDF needs
to be replaced by the general pair distribution function $g({\bf r},{\bf r}')$ from which the
modified 2-particle density ${\tilde R}({\bf r},{\bf r}')$
can be obtained through a multi-dimensional convolution. However, calculating the
two-point function $g({\bf r},{\bf r}')$ may be time-consuming.
Alternatively, the number fluctuations in a given sampling volume can be computed directly as in
the small system method~\cite{schnell11}. For particles that cross the surface of the sampling volume,
the volume fraction of the particle inside the volume can be computed analytically as a function of
the particle-surface distance, to ensure numerical efficiency.
Details on the generalization of the present theory to inhomogeneous systems will be given elsewhere.

\section*{Acknowledgments}
I thank Jean-Marc Simon for many stimulating discussions.

\appendix

\section{Why density fluctuations in a homogeneous system should be size-independent}
Here we argue that in a homogeneous system,
the relative particle number fluctuations~$\nu$ 
should
be size-independent to be physically meaningful, more precisely:
only if $\nu$ is size-independent can it be interpreted as a
thermodynamic response function.
We consider a homogeneous system at temperature $T$ and density
$\rho=N_0/V_0$, where $N_0$,$V_0\rightarrow\infty$.
Sampling is done in a finite subvolume~$V$.
The pressure of the subvolume, obtained by averaging over the $(N_0V_0T)$
ensemble and over all particles inside~$V$,
is in principle a function of $V$, $N_0$, $V_0$ and $T$,
but in a homogeneous system, $p$ is obviously independent of~$V$,
i.e. it is a function of $(N_0,V_0,T)$ alone. 
Assuming that $p$ depends on $V$ leads to a contradiction.
Indeed, when the total system is filled up with small identical sub-volumes~$V$,
the pressure in each subvolume must be the same as the macroscopic pressure, so
it cannot be size-dependent.
Now, by definition, the pressure function $p(N_0,V_0,T)$ completely determines
the isothermal compressibility as $1/\kappa_T = -V_0(\partial p/\partial V_0)_{TN_0}$.
Since $p$ is independent of $V$, so is $\kappa_T$.
The foregoing argument clearly shows that the compressibility of an {\it open}\/
sub-volume~$V$ of a {\it homogeneous}\/ system, is independent of~$V$.
So, if the compressibility is to be obtained from fluctuation theory,
the density fluctuations must be defined in such as way as to be
size-independent.
We note that the same reasoning can be done for the variable pair $(N_0,\mu)$
instead of ($V_0,p$). It follows that
the derivative of the chemical potential
$(\partial\log{N}/\partial\mu)_{TV}$, which is the response function
most directly related to the particle number fluctuations in an open
systeme~\cite{dawass19}, is also independent of the subvolume
size~$V$.

\end{document}